\documentclass{article}
\usepackage{gsirep}
\usepackage{graphicx}
\newcommand{\be}{\begin{equation}}
\newcommand{\ee}{\end{equation}}
\newcommand{\bea}{\begin{eqnarray}}
\newcommand{\eea}{\end{eqnarray}}
\title{Inhomogeneous freeze-out in heavy-ion collisions}
\author{A.~Dumitru$^a$, L.~Portugal$^{a,b}$ and D.~Zschiesche$^a$}
\affil{$^a$Institut f\"ur Theoretische Physik, Universit\"at Frankfurt
  a.M., Germany\\
$^b$Instituto de Fisica, Universidade Federal do Rio de Janeiro, Brazil}
\date{}
\begin{document}
\maketitle
\noindent 
Lattice QCD finds that a line of first-order phase transitions in
the $(\mu_B,T)$ plane ends in a critical point at
$T\approx160$~MeV, $\mu_B\approx360$~MeV~\cite{Fodor:2001pe}.
There is an ongoing experimental effort to detect that critical point
in heavy-ion collisions at high energies. It is hoped that by varying the
beam energy, for example, one can ``switch'' between the regimes of first-order
transition and cross over, respectively. 
If the particles decouple shortly after the expansion trajectory
crosses the line of first order transitions one may expect
a rather inhomogeneous (energy-) density distribution on the
freeze-out surface~\cite{Paech:2003fe} (similar, say, to the CMB photon
decoupling surface observed by WMAP~\cite{wmap}).
On the other hand, collisions at
very high energies ($\mu_B\simeq0$) may cool smoothly from high to low
$T$ and so pressure gradients tend to wash out density
inhomogeneities. Similarly, in the absence of phase-transition induced
non-equilibrium effects, the predicted initial-state density
inhomogeneities~\cite{iniflucs} should be strongly damped.

Here, we investigate the properties of an inhomogeneous fireball at (chemical)
decoupling. Note that if the scale of these inhomogeneities is much
smaller than the decoupling volume then
they can not be resolved individually, nor will they give rise to large
event-by-event fluctuations. Because of the nonlinear
dependence of the hadron densities on $T$ and $\mu_B$ they should nevertheless
reflect in the {\em event-averaged} abundances. Our goal is to
check whether the experimental data show any signs of
inhomogeneities on the freeze-out surface.

Perhaps the simplest possible {\it ansatz} is to employ the grand
canonical ensemble and to assume that the intensive
variables $T$ and $\mu_B$ are distributed according to a
Gaussian. This avoids reference to any
particular dynamical model for the formation and the distribution of density
perturbations on the freeze-out surface. Also, in this simple model
we do not need to
specify the probability distribution of volumes $V$.
Then, the average density of species $i$ is computed as
\bea
& &\overline{\rho}_i\; (\overline{T},\overline{\mu}_B, \delta T,\delta\mu_B)
= \\
& &\int\limits_0^\infty
 dT \; P(T;\overline{T},\delta T) 
\int\limits_{-\infty}^\infty 
d\mu_B \; P(\mu_B;
\overline{\mu}_B,\delta\mu_B)~\rho_i (T,\mu_B)~, \nonumber
\eea
with $\rho_i(T,\mu_B)$ the actual ``local'' density of species $i$, and
with $P(x; \overline{x},\delta x) \sim $ 
$\exp \left(-\frac{\left(x-\overline{x}\right)^2}{2\; \delta x^2} \right)$
the distribution of temperatures and chemical potentials on the
freeze-out surface.  
Feeding from (strong or weak) decays is included by replacing
%\be
$\overline{\rho}_i \rightarrow \overline{\rho}_i + B_{ij}\;
\overline{\rho}_j.$
%\ee
The implicit sum over $j\neq i$ runs over all unstable hadron species, with
$B_{ij}$ the branching ratio for the decay $j\to i$. For the present
analysis we computed the densities $\rho_i (T,\mu_B)$ in the ideal-gas
approximation. 

We perform a $\chi^2$ fit to both the midrapidity and $4 \pi$ data
 obtained by NA49 for central
Pb + Pb collisions at $\sqrt{s_{NN}} = 17.3$~GeV (compiled in
 \cite{Becattini:2003wp}). A similar analysis 
for lower and higher energies is underway.
Error estimates for the parameters 
(confidence intervals) are obtained from the projection of 
the regions in parameter space defined by $\chi^2 \le \chi^2_{min} +
1$ onto each axis. This corresponds to a
confidence level of $68.3 \%$ if the errors are normally distributed.

Table \ref{tab1} shows the resulting best fits with and
without finite widths of the $T$ and $\mu_B$ distributions.
\begin{table}[ht]
\begin{center}
\begin{small}
\begin{tabular}{|ccccc|} \hline
$\overline{T}$ & $\overline{\mu}_B$ & $\delta T$ & 
      $\delta\mu_B$ & $\chi^2/\mbox{dof}$ \\ \hline
& & SPS-158 (mid)& & \\
$155\pm 5$ & $200\pm 10$ & 0 & 0 & 40.4/8 \\
$105\pm 5$ & $230\pm 15$ & $35\pm 5$ & $80 \pm 40$ & 11.2/6 \\ \hline
& & SPS-158 ($4\pi$) & & \\
$145\pm 5$ & $210\pm 15$ & 0 & 0 & 40.0/11 \\
$100\pm 5$ & $260\pm 15$ & $30\pm5$ & $190\pm 35$ & 5.7/9 \\ \hline
\end{tabular}
\end{small}\end{center}
\vspace{-0.5cm}
\caption{\label{tab1}Fit results.
Lines with $\delta T=\delta\mu_B=0$
  correspond to a forced homogeneous fit. }
\end{table}
The fits improve (lower $\chi^2/\rm{dof}$) substantially
in both cases when $\delta T$, $\delta\mu_B$ are not forced to
zero. The inhomogeneous fits return significantly lower
mean temperature $\overline{T}$. However, this is  
{\em not} the ``mean'' temperature of the particles, which instead is
given by $\langle T \rangle_i = \int dT \;TP(T)
\int d\mu_B \; P(\mu_B) \;\rho_{i}
(T,\mu_B)/$ $\overline{\rho}_{i}$, cf.\ table \ref{tab2}.
\begin{table}[ht]
\begin{center}
\begin{small}
\begin{tabular}{|c|cccccc|} \hline
SPS 158  & $p$ & $\bar{p} $ & $K^+$ & $K^-$ & $\Omega$ & $\bar{\Omega}$ \\ \hline
$<T>$ [MeV] (mid)& 157 & 170 &  152 & 150 & 164 & 180 \\ 
$<\mu_B>$ [MeV] (mid)& 268 & 191 & 237 & 222 & 234 & 225 \\
\hline
$<T>$ [MeV] ($4\pi$)& 136 & 153 &  140 & 139 & 151 & 165 \\ 
$<\mu_B>$ [MeV] ($4\pi$)& 487 & 22 & 306 & 213 & 277 & 206 \\
\hline 
\end{tabular}
\end{small}\end{center}
\vspace{-0.5cm}
\caption{\label{tab2}Mean temperature and chemical potential 
of various particle species for the inhomogeneous freeze-out.}
\end{table}
We see that e.g.\ the bulk of the particles originates from 
different density and temperature 
regions than the corresponding anti-particles. 
Hence, our results suggest that the decoupling surface might
not be very well ``stirred''.

\end{document}